\documentclass[aps,preprint,nofootinbib]{revtex4}

\usepackage{graphicx}
\usepackage{epic}
\usepackage{eepic}
\usepackage{latexsym}

\newcommand{\eq}[1]{(\ref{#1})}
\newcommand{\be}{\begin{equation}}
\newcommand{\ee}{\end{equation}}
\newcommand{\bea}{\begin{eqnarray}}
\newcommand{\eea}{\end{eqnarray}}

\newcommand{\hs}[1]{\hspace{#1 mm}}

\def\a{\alpha}
\def\b{\beta}

\def\d{\delta}
\def\D{\Delta}
\def\e{\epsilon}

\def\fr{\frac}

\def\l{\lambda}

\def\m{\mu}
\def\n{\nu}

\def\r{\rho}
\def\s{\sigma}
\def\S{\Sigma}
\def\t{\tau}

\def\O{\Omega}

\def\o{\omega}

\def\del{\partial}

\def\nn{\nonumber}

\begin{document}

\title{Volume Stabilization and Acceleration \\in Brane Gas Cosmology}

\author{Ali Kaya}
\email[]{kaya@gursey.gov.tr}

\affiliation{Feza G\"{u}rsey Institute,\\
\c{C}engelk\"{o}y, 81220, \.Istanbul, Turkey} 

\date{\today}

\begin{abstract}
We investigate toy cosmological models in
$(1+m+p)$-dimensions with gas of $p$-branes wrapping over $p$-compact
dimensions. In addition to winding modes, we consider the effects of
momentum modes corresponding to small vibrations of branes and find that the
extra dimensions are dynamically stabilized while the others
expand. Adding matter, the compact volume may grow slowly depending on
the equation of state. We also obtain  solutions with winding and
momentum modes where the observed space undergoes accelerated expansion.     
\end{abstract}


\maketitle

\section{Introduction}

String/M theory predicts six/seven extra dimensions which are
presumably compact and very small compared to the observed three  
dimensions. To find out how extra dimensions are compactified is a
major problem in the theory waiting for a solution. This is important
in determining the vacuum structure and the low energy content which
may then be compared with experimental findings to test
validity of string/M theory. It is known that phenomenologically
viable models can be obtained in Calabi-Yau or G2 manifold
compactifications.      

Cosmology, on the other hand, is another framework where ideas in
string/M theory may find observational support. We know that the
perceived universe was initially small and grew to its size today
after various cosmological eras. If extra dimensions exist they should
have an effect on the cosmological evolution. However, since standard
cosmology offers a successful scenario which remains plausible close
to the temperatures around Tev scale, the effects of extra dimensions
should be negligible from that time till today. For example, the
theory of big-bang nucleosynthesis together with the observational
abundances of light elements provide a limit to the change in the size
of internal space following the production of these
elements. Therefore, it is natural to assume that the extra dimensions
were already stabilized around the epoch of nucleosynthesis.   

A mechanism for stabilization based on the cosmological impact of
string winding and momentum modes was proposed in \cite{bv} (see also
\cite{eski1} and \cite{eski2} for earlier work on string winding modes
in cosmology). In the same paper, it is suggested that the initial
big-bang singularity can be resolved by T-duality invariance of string
theory and the dimensionality of observed space-time can be explained
via the annihilation of winding strings leading decompactification of
three dimensions. The proposal of \cite{bv} was generalized in
\cite{abe} to include other branes in string/M theory and it is now 
an alternative approach named as brane gas cosmology (for
recent work see e.g. \cite{bg1,bg2,bg3,bg4,bg5,bg6,bg7,bg8,biz1,biz2,
bgb1,bcamp,bgb2,bgb3,sb1,sb2,sb3,sb4}). 
In this scenario, strings dominate the the late time evolution and
thus the stabilization of extra dimensions {\it today} should be
achieved by string winding and momentum modes. This is quantified in
the papers \cite{sb1,sb2,sb3,sb4} where solutions to Einstein and
dilaton gravity equations with  stabilized  internal dimensions are
presented.   

However, this mechanism seems to work for toroidal
compactifications which are not phenomenologically viable. For
example, it is  not clear how strings may stabilize {\it volume} of a
Calabi-Yau or a G2 manifold since the topology is not equal to the 
product of one-dimensional cycles (this is much more evident for a
sphere compactification since there cannot be stable winding strings
at all). Therefore, to fix the moduli related to higher dimensional
non-trivial (i.e. non-toroidal) topological cycles, it seems that the
higher dimensional branes are also needed to play a cosmological role
at late times.    

Motivated by this argument, in this paper we consider a {\it
toy} cosmological model in $(1+m+p)$-dimensions with a gas of $p$-branes
wrapping over the $p$-dimensional compact space, and try to determine
whether the compact volume is stabilized. We should emphasize that such
a model does not straightforwardly lead to a realistic scenario. Apart from
the fact that all other possible (stringy) excitations are ignored,
the existence of a gas of 6 or 7-branes is also dubious
(note that in a realistic model one should set $m=3$ and $p=6,7$)
since in the very  early universe the higher dimensional branes are
expected to annihilate leaving a gas of lower dimensional branes at
late times. In any case, the results obtained in such toy models
are still expected to be useful in brane gas cosmology.    

As in the original proposal of \cite{bv}, one would anticipate a
balance between  {\it brane} winding and momentum pressures for
stabilization. The energy momentum tensor for
winding part was already determined in the cosmological context (see
e.g. \cite{bg5} or \cite{biz2}). Since we do not know how to quantize
$p$-branes for $p>1$, the exact spectrum of the momentum modes is not
known. However, in a semiclassical approximation one may consider small
fluctuations around a rigid wrapped $p$-brane and from this
spectrum an energy  momentum tensor for momentum modes can be
determined. In section \ref{II}, we will work out this problem and
study the issue of stability in the context of Einstein gravity. We
find that, like strings, gas of $p$-branes with $p>1$ can also
dynamically stabilize extra dimensions. 

Based on recent observations one may claim that any potentially
realistic cosmological model should include a theory for accelerated
expansion. It is well known that inflationary paradigm is successful
in resolving the basic shortcomings of standard cosmology. On the
other hand, there are strong observational indications that the
universe undergoes an accelerated expansion today. In brane gas
cosmology, a way of realizing power-law inflation driven by a gas of
domain walls was proposed in \cite{inf}. In section \ref{III}, we also
obtain accelerating solutions with brane winding and momentum modes
(the background of winding modes generalizes the power-law metric
constructed in \cite{biz1}). As we will discuss, depending on the
initial conditions, one encounters two different behavior: either a
short period of acceleration with a small number (like 0.7) of
e-foldings or an infinite accelerated  expansion ending at a
singularity at finite proper time.    

\section{Stabilization}\label{II}

Consider a $(1+m+p)$-dimensional space-time with the metric
\be
ds^2=-dt^2+dx^idx^i+R^2\, d\S_p^2 \label{stmet},
\ee
where $i,j=1,..,m$, $\S_p$ is a $p$-dimensional compact manifold, $R$
is a constant scale factor and the metric on $\S_p$ can be written as $(a,b
=1,..,p)$, 
\be 
d\S_p^2\,=\,g_{ab}(y)\,dy^a dy^b.
\ee
In this paper we take $g_{ab}$ to be Ricci flat. For other cases, we
expect the internal curvature terms to become important at late times
as in \cite{bgb1}. In the physical gauge (and for small curvatures)
the action for a $p$-brane wrapping over $\S_p$ can be written as    
\bea
S_p&=&-T_p\int \,d^{p+1}\s\,\sqrt{-\gamma},\nn\\
&=& -T_p\int d^{p+1}\s\,\sqrt{-\gamma_0}\,-\,\fr{T_p}{2}\,\int
d^{p+1}\s\,\sqrt{-\gamma_0}\,\gamma_0^{\a\b}\,\del_\a x^i\,\del_\b
x^i\,+\,{\mathcal O}(x^4),\label{qa} 
\eea 
where $T_p$ is the brane tension, $\s^{\a}=(t,y^a)$ are the local brane
coordinates, $\gamma^0_{\a\b}$ is the zeroth order induced metric given by 
\be
\gamma^0_{\a\b}\,d\s^\a \,d\s^\b\,=\,-dt^2\,+\,R^2\,d\S_p^2,
\ee
and in the last line in \eq{qa} we expand the action to the second order in
the transverse fluctuation fields $x^i(\s^\a)$. 

The first term in \eq{qa} corresponds to the winding modes which have
the energy spectrum  
\be
E \,=\,(n\,T_p \,\O_p)\,R^p, \label{we}
\ee
where $n$ is winding number and $\O_p$ is the volume of $\S_p$. From
\eq{we}, the total pressure in the internal space can be found as 
\be
P\,=\,-p\,(n\,T_p \,\O_p)\,R^p,\label{wp}
\ee
which turns out to be negative, as expected. 

On the other hand, \eq{qa} yields the following field equation for $x_i$:  
\be
\ddot{x}_i-\fr{1}{R^2}\,\nabla^2\,x_i=0, \label{x}
\ee
where the dot denotes derivative with respect to $t$ and $\nabla^2$ is
the Laplacian on $\S_p$. Expanding in terms of the harmonics  on
$\S_p$, the spectrum of {\it momentum} modes (i.e. the
eigenfrequencies $e^{i\o t}$) can be obtained as 
$\o^2=\l_n^2/R^2$, where $-\l_n^2$ are the eigenvalues of the
Laplacian. Thus the energy for the $n$'th momentum mode is given by
(we choose $\l_n>0$) 
\be
E\,=\,\fr{\l_n}{R} \label{me}, 
\ee
and the corresponding total pressure on $\S_p$ is 
\be
P\,=\,\fr{\l_n}{R}, \label{mp}
\ee
which is positive contrary to \eq{wp}. Although \eq{me} and \eq{mp}
depend on the undetermined eigenvalues $\l_n$, one does not need
these numbers to proceed. On the other hand,  
$1/R$ dependence in \eq{mp} can also be deduced from the uncertainty
principle $\D y\,\D P\sim 1$ where a compact direction $y$ in $\S_p$
approximately satisfies $\D y\sim R$. This suggests that \eq{me} and
\eq{mp} should also hold for other possible fields propagating on the
world-volume such as U(1) gauge fields on D-branes or self-dual
3-form of M5 brane. Gauging away the longitudinal polarizations, the
linearized transverse modes of these brane fields are expected to obey
\eq{x}.     

In a cosmological setting, \eq{stmet} should be replaced with
\be\label{met1}
ds^2=-e^{2A}dt^2+e^{2B}dx^idx^i+e^{2C}\,d\S_p^2,
\ee
where the metric functions $A$, $B$ and $C$ depend only on time
$t$. The scale factors for the observed and the internal
spaces are defined as 
\be \label{robrin}
R_{ob}=e^{B},\hs{8}R_{in}=e^{C}.
\ee
Note that $t$ is {\it not} the proper time and we have not yet fixed the
corresponding reparametrization invariance. 

From \eq{we}-\eq{mp}, the energy momentum tensor
for the brane winding and momentum modes can be obtained by
calculating the corresponding {\it densities} (i.e. by dividing with
the total spatial volume $V=e^{mB+pC}$) which gives   
\bea
T_{\hat{t}\hat{t}}&=&T_w\, e^{-mB}\,+\,T_m\,e^{-mB-(p+1)C},\nonumber\\
T_{\hat{i}\hat{j}}&=&0,\label{enmom1}\\
T_{\hat{a}\hat{b}}&=&-T_w \,e^{-mB}\,\d_{ab}\,+\,\fr{T_m}{p}\,
e^{-mB-(p+1)C}\,\d_{ab},\nonumber 
\eea
where  $T_w$ and $T_m$ are constants (which can in principle be fixed
by the parameters of the underlying fundamental theory and
thermodynamics) and the hatted indices refer to the tangent space. It
is easy to verify that $\nabla_\m T^{\m\n}=0$. In driving \eq{enmom1},
the total pressures given in \eq{wp} and \eq{mp} are assumed to be
distributed isotropically on the tangent space of $\S_p$ since 
the momentum modes are point-like excitations (note that we
are doing field theory on the world-volume) and thus locally they
would only see the flat, tangent space. However, to justify this
assumption more rigorously one may additionally require that $\S_p$ is
not highly curved and thus has a ``smooth'' shape with no sharp
corners or handles, which is also necessary for the validity of the
brane action and the small field approximation used in \eq{qa}. Let us
note that the contribution of the winding part in \eq{enmom1} agrees
with the expressions derived in \cite{biz1, biz2} which are obtained
in a different way by coupling the brane action to the gravity
action. This agreement further supports the above considerations.   

In general, if one has the metric
\be\label{met2}
ds^2=-e^{2A}dt^2+\sum_{i}e^{2B_i}dx^idx^i,
\ee
and the energy momentum tensor 
\be
T_{\hat{\mu}\hat{\nu}}=\textrm{diag}\,(\rho,p_{\hat{i}}),\label{energy1}
\ee
where $p_{\hat{i}}=\omega_{i}\,\rho$ with constant $\omega_i$, the
conservation equation $\nabla_\mu T^{\mu\nu}=0$ gives
\be\label{omega}
\rho=\rho_0\,\,\exp\left[-\sum_{i}(1+\omega_i)B_i\right]\, ,
\ee
where $\rho_0$ is a constant. Comparing to \eq{enmom1}, we find that
the energy momentum tensor for winding and momentum modes can be
written as \eq{energy1} where  for winding modes 
\be
\omega_i=\left\{\begin{array}{ll} 
-1:\,\,\,\textrm{brane direction},\\
\,\,0\hs{2}:\,\,\,\textrm{transverse direction},\end{array}\right.
\label{w}
\ee
and for momentum modes
\be
\omega_i=\left\{\begin{array}{ll} 1/p:\,\,\,\textrm{brane direction},\\
\,\,\,\,0\,\,:\,\,\,\,\textrm{transverse direction},\end{array}\right.
\label{m}
\ee
which corresponds to radiation confined in the compact 
space.\footnote{The energy-momentum tensor for \eq{w} or \eq{m} can
  also be
  viewed to correspond to a perfect fluid in $(1+m+p)$-dimensions. 
  Certain aspects of higher dimensional cosmological
  models with general multi-component perfect fluid matter were
  studied in the past, see e.g. \cite{f1,f2,f3,f4}.}
With the above formulas it is possible to determine the energy
momentum tensor for any given brane configuration and study the
resulting cosmological evolution. 

Using \eq{enmom1} and imposing the gauge $A=m\,B+p\,C$ in \eq{met1},
which fixes the $t$-reparametrization invariance, Einstein equations
can be written as follows: 
\bea
&&\ddot{A}-\dot{A}^2 +m\,\dot{B}^2+p\,\dot{C}^2=
-\fr{m-2}{m+p-1}\,T_w\,e^{mB+2pC}-T_m\,e^{mB+(p-1)C},\nn\\
&&\ddot{B}=\frac{p+1}{m+p-1}\,T_w\,e^{mB+2pC},\label{e1}\\
&&\ddot{C}=-\frac{m-2}{m+p-1}\,T_w\,e^{mB+2pC}+\fr{T_m}{p}\,e^{mB+(p-1)C},
\nn
\eea
where the dot denotes derivative with respect to $t$ and the
gravitational constant is set to one.  

We could not obtain the most general solution of \eq{e1}. However, it
is easy to verify that the functions 
\be\label{19}
A=-2\ln (\a\,t)+p\,C_0, \hs{5}B=\fr{-2}{m}\ln (\a \,t),\hs{5} C=C_0,
\ee
give a solution where $\a$ is a constant, which is
uniquely\footnote{The field equations give   
\bea \a^2=\fr{m(p+1)T_m}{2p(m-2)}\,(R_{in}^0)^{p-1}. \nn\eea }
fixed by field equations, and $C_0$ can be calculated from the third
equation as 
\be\label{sd}
e^{C_0}=
R_{in}^{0}=\left[\fr{T_m\,(m+p-1)}{T_w\,(m-2)\,p}\right]^{1/(p+1)}. 
\ee
Introducing the proper time coordinate 
\be\label{tau}
d\tau=-e^{A}\, dt,
\ee
we get 
\be\label{stabmet}
ds^2\,=\,-d\t^2\,+\,(\a\,\t)^{4/m}\,dx^idx^i\,+\,(R_{in}^0)^2\,d\S_p^2,
\ee
where the internal dimensions are stabilized and the power-law of the
observed space is exactly the same as the one for pressureless matter
in standard cosmology. Here, $R_{in}^0$ can be viewed as the self-dual
radius corresponding to brane winding and momentum
modes. Technically, $C$ is allowed to be a constant since in the
third equation in \eq{e1}, $\exp(mB)$ terms can be factored out and
there are both negative and positive contributions to $\ddot{C}$. 

The metric \eq{stabmet} corresponds to the initial data
where the internal dimensions begin at the self-dual radius
with zero velocity. For more general initial conditions one may integrate
\eq{e1} numerically. The first equation (when combined with the other 
two) in \eq{e1} gives a constraint on the initial data: 
\be
m(m-1)\left(\fr{dB}{d\t}\right)^2+p(p-1)\left(\fr{dC}{d\t}\right)^2
+2mp\left(\fr{dB}{d\t}\right)\left(\fr{dC}{d\t}\right)
-2T_we^{-mB}-2T_me^{-mB-(p+1)C}=0,\label{ini}
\ee  
where $\t$ is the proper time defined in \eq{tau}. Fixing $B(\t)$, $C(\t)$ 
and one of the velocities\footnote{In our numerical integrations, we
  first fix   $dC/d\tau$ and solve $dB/d\tau$ from \eq{ini}.}
$dB/d\tau$ or $dC/d\tau$ at some $\t=\t_0$,
\eq{ini} gives {\it two real} roots for the undetermined initial
velocity. It is easy to see that one of the roots should 
have the opposite sign compared to the other fixed initial
velocity. 

\begin{figure}
\centerline{\includegraphics[width=10.0cm]{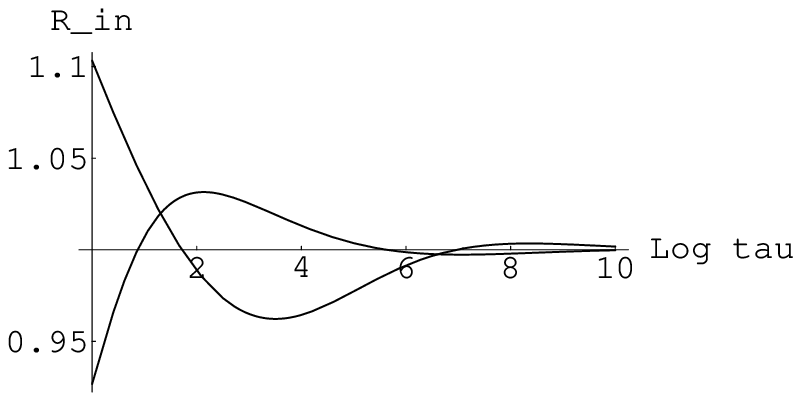}}
\caption{\label{I} The graphs of $R_{in}(\t)$ for two different
  initial data set obeying (i). For convenience we take $m=3$, $p=6$,
  $T_w=8$ and $T_m=6$ so that the self-dual radius \eq{sd} is equal to
  1. The initial conditions are: $R_{ob}(0)=1$,
  $R_{in}(0)=0.7$, $dR_{ob}/d\t(0)=3.9761$, $dR_{in}/d\t(0)=0.3$ and
  $R_{ob}(0)=1$, $R_{in}(0)=1.3$, $dR_{ob}/d\t(0)=3.65135$,
  $dR_{in}/d\t(0)=-0.7.$ Here, $dR_{ob}/d\t(0)$ is solved from
  \eq{ini} given other conditions.} 
\end{figure}

After several numerical integrations of \eq{e1} for
different initial conditions and for $(m,p)=(3,6)$  (these $(m,p)$
values are suggested by string theory, however we expect the same
conclusions to hold for any set with $m>2$) we observe the following
generic behavior: 

(i) For one of the roots, $\exp(C)=R_{in}$ performs damped oscillations
around the self-dual radius given in \eq{sd} and the observed space
expands with negative acceleration close to the power-law in
\eq{stabmet}. As the extra dimensions are stabilized asymptotically, 
the solution becomes \eq{stabmet}. In figure \ref{I}, we plot two
such numerical runs for $R_{in}$. As in \cite{sb1}, the evolution of the
large dimensions is responsible for these oscillations to be damped.   

(ii) For the other root, the numerical integration signals a singularity
at finite proper time, where the observed and the internal spaces 
have the opposite behavior near the singularity i.e. if one expands
the other contracts. Below, we will try to explain this case
analytically.  

(iii) If the initial values of $dB/d\tau$ and $dC/d\tau$ are  positive
(i.e. if both spaces are expanding at the beginning) then we
observe the stabilization mentioned in (i). For the other cases both (i)
and (ii) are possible.

It is not surprising that for some initial conditions the metric  
ends with a singularity at finite proper time. We do not expect to obtain a
completely smooth cosmological solution with the ansatz \eq{met1}.
Contrary, at least one singular fixed point in the past or in the
future is anticipated. Therefore, for the alternatives (i) and (ii) the
following interpretation can be given: in the phase space one of the roots
selects a trajectory  which evolves ``forwards in time'' and gives the 
smooth stabilized solution, and the other root corresponds to a path
where the space-time evolves ``backwards in time'' and hits the
initial singularity.   

Actually, to define the arrow of time in such solutions an extra
(observational) input is needed. For instance, we use the
Robertson-Walker metric starting from the big-bang singularity since
we observe that the universe at the moment is expanding. If we would
measure a contraction rather than an expansion, the same metric should be
considered with $\t\to-\t$, i.e. with a future big-crunch
singularity instead. In our case, there is no such avaliable 
observational input. However, it seems reasonable to assume that
\eq{e1} holds some time after the big-bang explosion and thus one should take
all dimensions to be expanding initially in \eq{e1}. From (iii) we
then see that the corresponding solution asymptotes to \eq{stabmet}
and the stabilization can be achieved. On the other hand, we should also
bear in mind that there are physically acceptable initial
data, such as an expanding observed and contracting internal
dimensions at the beginning, where the subsequent evolution ends 
with a naked singularity at finite proper time. 

To have a better understanding of this singular behavior, let us
analyze \eq{e1} more carefully. Comparing the contributions of the
winding and momentum modes, we see that when $\exp[(p+1)C]\gg T_m/T_w$
the winding modes and when  $\exp[(p+1)C]\ll T_m/T_w$ the momentum
modes dominate the evolution. So, it would be useful to solve \eq{e1}
in these two limits. 

To consider the effects of winding modes alone we set $T_m=0$. Then the last
two equations in \eq{e1} implies
\be\label{c}
C\,=\,-\fr{m-2}{p+1}B\,+ a\,t,
\ee
where $a$ is a constant. Using this relation in \eq{e1}, one gets a
second order differential equation for $B$ which can be solved
(up to an irrelevant additive constant depending on $T_w$) as
\be\label{b}
B=-\fr{2(p+1)}{m-mp+4p}\,\ln(\sinh(b\,t))-\fr{2p(p+1)}{m-mp+4p}\,a\,t
\ee
where $b$ is another constant. Using \eq{c}, \eq{b} and the gauge
$A=m\,B+p\,C$ in the first equation in \eq{e1} one finds  
\be\label{ab}
a^2=\fr{4(m+p-1)}{mp(p+1)^2}\,b^2,
\ee
and this completes the solution. In terms of the proper time coordinate
\eq{tau}, we see that as $t\to0$, $\t\to\infty$ and as $t\to\infty$,
$\t\to0$. In these two limits the solution becomes
\be
ds_w^2=\begin{cases}{
-d\t^2+(\t)^{2\,r_1}\,dx^i\,dx^i\,+\,
(\t)^{2\,r_2}\,d\S_p^2\hs{8}:\textrm{as}\hs{3}\t\to 0,\cr\cr
 -d\t^2+(\t)^{\fr{4}{m}}\,dx^i\,dx^i\,+\,
(\t)^{-\fr{4(m-2)}{m(p+1)}}\,d\S_p^2\hs{3}:\textrm{as}\hs{3}\t\to
  \infty, }\end{cases}\label{lim1}
\ee
where the powers $r_1$ and $r_2$ are given by
\bea
r_1&=&\fr{m\,+\,\e\,\sqrt{m\,p\,(m+p-1)}}{m(m+p)}\label{r1}, \\
r_2&=&\fr{p\,-\,\e\,\sqrt{m\,p\,(m+p-1)}}{p(m+p)}\label{r2}.
\eea
Here, $\e=\pm 1$ corresponds to the positive and the negative roots of
\eq{ab}, respectively, and in \eq{lim1} we ignore constants multiplying $\t$ in
parenthesis which can be set to 1 by scalings of $x$ and $y$
coordinates. We see that the full metric smoothly interpolates between these
two limiting geometries.   

Asymptotically, as $\t\to\infty$ the solution approaches the power-law
metric given in \cite{biz1} (see eq. (11) in that paper). On the other hand, as
$\t\to0$ one finds a vacuum Kasner metric since the powers $r_1$ and
$r_2$ obey 
\bea
&&m\, r_1\,+\,p \,r_2=1,\\
&&m\, r_1^2\,+p\, r_2^2=1.
\eea
Indeed, using \eq{c} and \eq{b} it is easy to see that {\it all} the source
terms on the right hand side of \eq{e1} vanish as $t\to\infty$ (and
thus $\t\to0$) so ending up with a vacuum solution is not surprising. 
This also shows that the same vacuum metric solves \eq{e1} with
$T_m\not =0$ in the $\t\to0$ limit. 

To see the effects of momentum modes alone let us now take $T_w=0$. In this
case the second equation in \eq{e1} gives
\be\label{m1}
B\,=\, a\,t,
\ee
where $a$ is a constant, and the third equation can be solved for $C$
(up to an irrelevant additive constant depending on $T_m$) as 
\be\label{m2}
C=-\fr{2}{p-1}\,\ln(\sinh(b\,t))\,-\,\fr{m}{p-1}\,a\,t.
\ee
The unknown function $A$ can be fixed from the gauge condition
$A=m\,B+p\,C$, and the first equation in \eq{e1} imposes 
\be\label{ab2}
a^2\,=\,\fr{4p}{m(m+p-1)}\,b^2.
\ee
In terms of the proper time \eq{tau}, we find that as $t\to (0,\infty)$,
$\tau\to(\infty,0)$, respectively, and asymptotically the solution becomes
\be
ds_m^2=\begin{cases}{
-d\t^2+(\t)^{2\,r_1}\,dx^i\,dx^i\,+\,
(\t)^{2\,r_2}\,d\S_p^2\hs{3}:\textrm{as}\hs{3}\t\to 0,\cr\cr
 -d\t^2\,+\,dx^i\,dx^i\,+\,
(\t)^{\fr{4}{p+1}}\,d\S_p^2\hs{13}:\textrm{as}\hs{3}\t\to
  \infty, }\end{cases}\label{lim2}
\ee
where $r_1$ and $r_2$ are given by \eq{r1} and \eq{r2} in which 
$\e=+1$ corresponds to the negative root of \eq{ab2} while $\e=-1$
corresponds to the positive root. As $\t\to0$, one again encounters
the {\it same} vacuum Kasner metric.  

Returning back the situation where both $T_w$ and $T_m$ are non-zero,
we see from \eq{lim1} and \eq{lim2} that as $\t\to0$ the solution
should approach the Kasner background with the powers $r_1$ and
$r_2$. On the other hand,  the field equations show  
that as $\t\to\infty$ the metrics given in \eq{lim1} and \eq{lim2}
should be modified i.e. in \eq{lim1} momentum modes can no longer be
ignored and in \eq{lim2} winding modes should be taken into
account. In either case the neglected contribution becomes dominant in
time, alters the 
behavior of the extra dimensions and serves for stabilization. This
suggests that asymptotically the most general solution of \eq{e1}
should become     
\be
ds^2=\begin{cases}{
-d\t^2+(\t)^{2\,r_1}\,dx^i\,dx^i\,+\,
(\t)^{2\,r_2}\,d\S_p^2\hs{3}:\textrm{as}\hs{3}\t\to 0,\cr\cr
 -d\t^2\,+\,(\t)^{\fr{4}{m}}\,dx^i\,dx^i\,+\,
(R_{in}^0)^2\,d\S_p^2\hs{3}:\textrm{as}\hs{3}\t\to
  \infty.}\end{cases}\label{full}
\ee
The numerical integrations of \eq{e1}, which are summarized in
(i)-(iii) above, also support this result. We already
mentioned in (i) that for some initial conditions the metric
approaches to \eq{stabmet}. The singularity corresponding to
other initial data indicated in (ii) occurs at $\t=0$ in
\eq{full}. Recall from (ii) that in reaching the singularity the
observed and the internal spaces have the opposite behavior. From 
\eq{r1} and \eq{r2}, we see that $r_1$ and $r_2$ have the opposite
signs and thus \eq{full} also explains this numerical result. As
discussed above, such initial conditions are associated with
solution evolving ``backwards in time''.     

Having studied the stabilization problem with brane winding and
momentum modes, let us now try to include the effects of ordinary
matter. These should be taken into account in determining the late
time behavior since in the universe today the existence of radiation
and pressureless dust is known. We take the following energy momentum
tensor for matter   
\be\label{emmat}
T_{\hat{\m}\hat{\n}}=\textrm{diag}(\,\r,\,p_{\hat{i}},\,p_{\hat{a}}\,),
\ee  
where 
\be
p_{\hat{i}}=\o\,\r,\hs{10}p_{\hat{a}}=\n\,\r,
\ee
and ($\o,\n$) are constants. Adding \eq{emmat}, the Einstein
equations \eq{e1} are modified as follows (we again impose the gauge
$A=m\,B+p\,C$):  
\bea
&&\ddot{A}-\dot{A}^2 +m\,\dot{B}^2+p\,\dot{C}^2=
-\fr{m-2}{m+p-1}\,T_wF_w-T_mF_m\,+\,
\r_0\left[\fr{2-m(\o+1)-p(\n+1)}{m+p-1}\right]F_\r
\nn\\
&&\ddot{B}=\frac{p+1}{m+p-1}\,T_w\,F_w\,+\,\fr{\r_0}{m+p-1}
\left[1+(p-1)\o-p\n \right] F_\r,\label{e2}\\
&&\ddot{C}=-\frac{m-2}{m+p-1}\,T_w\,F_w+\fr{T_m}{p}\,F_m\,
+\,\fr{\r_0}{m+p-1}\left[1+(m-1)\n-m\o \right] F_\r,
\nn
\eea
where
\bea
F_w&=&e^{mB+2pC},\nn\\
F_m&=&e^{mB+(p-1)C},\label{F}\\
F_\r&=&e^{(1-\o)mB+(1-\n)pC}.\nn
\eea
It is difficult to solve \eq{e2} in its most general form. However,
for $\o=0$, there is again a special solution which can be written in
the proper time coordinate as: 
\be\label{stabmet2}
ds^2\,=\,-d\t^2\,+\,(\a\,\t)^{4/m}\,dx^idx^i\,+\,(R_{in}^0)^2\,d\S_p^2,
\ee
where the self-dual radius $R_{in}^0$ obeys 
\be\label{sd2}
\fr{m-2}{m+p-1}\,T_w\,(R_{in}^0)^{p+1}-\fr{1+(m-1)\n}{m+p-1}\,\r_0\,
(R_{in}^0)^{1-\n p} -\fr{T_m}{p}=0,
\ee
and $\a$ is a fixed\footnote{The field equations impose 
\bea \a^2=\fr{m(p+1)T_m}{2p(m-2)}\,(R_{in}^0)^{p-1}
+\fr{m(\n+1)\r_0}{2(m-2)}\,(R_{in}^0)^{p-\n p} . \nn\eea}, positive constant. 

\begin{figure}
\centerline{\includegraphics[width=7.0cm]{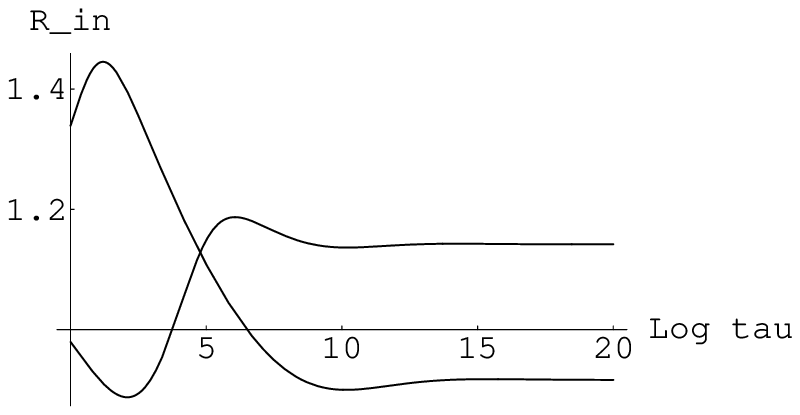}\hs{10}
\includegraphics[width=7.0cm]{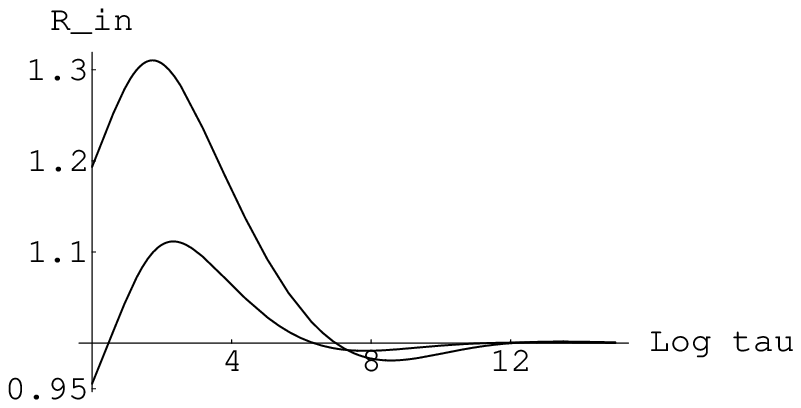}}
\centerline{(a)\hs{80}(b)}
\centerline{\includegraphics[width=7.0cm]{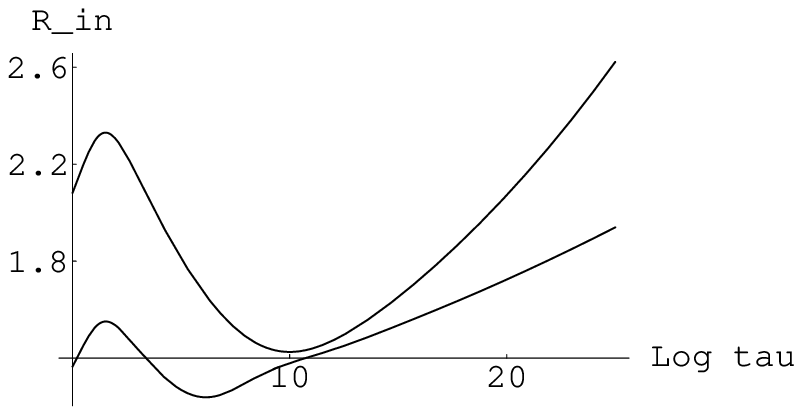}}
\centerline{(c)}
\caption{\label{22} The graphs of $R_{in}(\t)$ for different values of
  $\o$ and $\n$. For convenience we take $m=3$, $p=6$,
  $T_w=8$, $T_m=6$,  $R_{ob}(0)=2$. The initial conditions are: \newline
(a) $\left[\o=0\right]$: $\n=-0.9$, $R_{in}(0)=1.1$,
  $dR_{ob}/d\t(0)=-1.41365$, $dR_{in}/d\t(0)=1$;\newline    
    $\n=0.5$, $R_{in}(0)=1.1$, $dR_{ob}/d\t(0)=6.80001$,
  $dR_{in}/d\t(0)=-0.7$,\newline 
(b) $\left[\o=0.5\right]$: $\n=-0.6$, $R_{in}(0)=1.01$,
  $dR_{ob}/d\t(0)=1.10393$, $dR_{in}/d\t(0)=0.1$;\newline    
    $\n=0.3$, $R_{in}(0)=0.8$, $dR_{ob}/d\t(0)=1.07248$,
  $dR_{in}/d\t(0)=0.3$,\newline 
(c) $\left[\o=-0.1\right]$: $\n=-0.3$, $R_{in}(0)=1.7$,
  $dR_{ob}/d\t(0)=-0.870317$, $dR_{in}/d\t(0)=1$;\newline     
    $\n=0.4$, $R_{in}(0)=1.1$, $dR_{ob}/d\t(0)=-1.26295$,
  $dR_{in}/d\t(0)=0.9.$\newline 
Here, $dR_{ob}/d\t(0)$ is solved from the constraint on initial data. 
} 
\end{figure}

The metric \eq{stabmet2} evolves from the initial data where the
internal space starts at the self-dual radius with zero velocity. For
more general initial conditions, we numerically integrate\footnote{As
  before, we first fix all but the velocity of the observed space at
  $\t=0$ and solve $dB(\t)/d\t$ from the constraint equation which
  gives two real roots. In the following, we only consider
  one of the roots which yields an evolution ``forwards in time''.} 
\eq{e2} and our findings can be summarized as follows:

(a) $\o=0$: We find that the internal dimensions are stabilized after
some damped oscillations around the self-dual radius determined in
\eq{sd2} and the observed space expands. Asymptotically, the solution
approaches to \eq{stabmet2}. In figure \ref{22}(a), two such numerical
integrations for $R_{in}(\t)$ are plotted. Note that, from \eq{sd2}
the final size of the extra dimensions depends on $\n$.    

(b) $\o>0$: We find that the effect of ordinary matter on the
evolution is negligible at late times, i.e. brane winding
and momentum modes dominate the cosmology. The solution asymptotically
becomes \eq{stabmet} where the self-dual radius given by \eq{sd}
does not depend on $\n$. In figure \ref{22}(b), we plot two
numerical runs for $R_{in}(\t)$. The initial conditions
are chosen so that $R_{in}^0=1$.     

(c) $\o<0$: In this case, the late time evolution is dominated by 
matter and the stabilization {\it cannot} be achieved. Two numerical runs
for $R_{in}(\t)$  are plotted in figure \ref{22}(c). At early times,
the oscillations caused by winding and momentum modes can be seen in
the figure. However, matter finally causes the internal space to grow
at late times.    

From \eq{e2}, we see that the contributions of matter, the winding and
the momentum modes are proportional to the functions $F_\r$, $F_w$ and
$F_m$ given in \eq{F}, respectively. Assuming that the size of the
observed space becomes much larger than the size of the extra
dimensions in time\footnote{This is expected since the winding
modes resist the expansion of the extra dimensions, however all modes
contribute to the growth of the observed space.}, one sees that $e^B$
factors in \eq{F} dictate a hierarchy between $F$-functions.
This explains the generic behavior we encountered in (a)-(c):
compared to $F_w$ and $F_m$, the function $F_\r$ gets stronger for
$\o<0$ and it becomes negligible for $\o>0$. In the former case it is
clear that the stabilization mechanism does not work. When $\o=0$ all
functions have the same strength and stabilization can be achieved due
to the presence of positive and negative contributions to $\ddot{C}$
in \eq{e2}, where the self-dual radius now depends on the equation of
state. These considerations suggest that if the acceleration
of the universe is due to some form of matter with $\o<0$ (i.e. dark
energy), then the extra dimensions cannot be stabilized by brane
winding and momentum modes. We expect the same conclusion to hold for
the strings studied in \cite{sb1}. However, depending on the actual
values of the parameters in these equations, the change in the size of
the extra dimensions may be slow enough for us not to recognize it in
our time scale.        
  
\section{Acceleration}\label{III}

In this section, we show that some of the solutions constructed above
with  $T_m=0$ or $T_w=0$ have periods during which the observed space
undergoes accelerated expansion. Here, we do not intend to propose a
realistic model for inflation or current acceleration
of the universe, but simply note an alternative way of obtaining
acceleration in brane gas cosmology. Motivated by string/M
theory as the natural fundamental framework, in this section we also
set $m=3$ and $p=6$ for convenience.      

\begin{figure}
\centerline{\includegraphics[width=8.0cm]{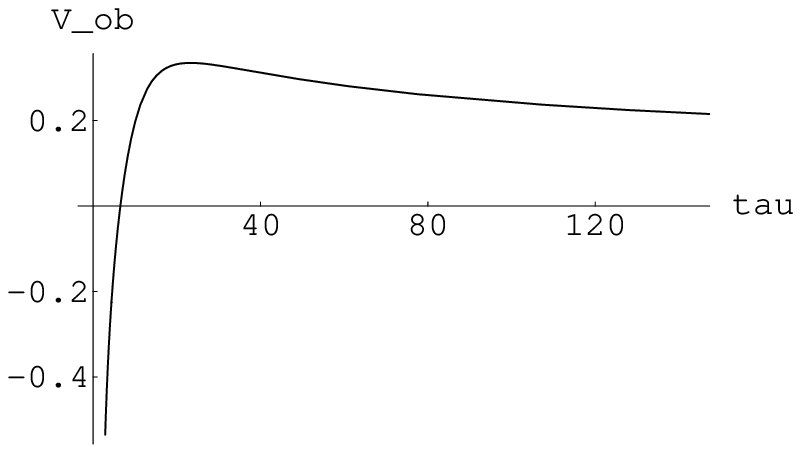}\hs{6} 
\includegraphics[width=8.0cm]{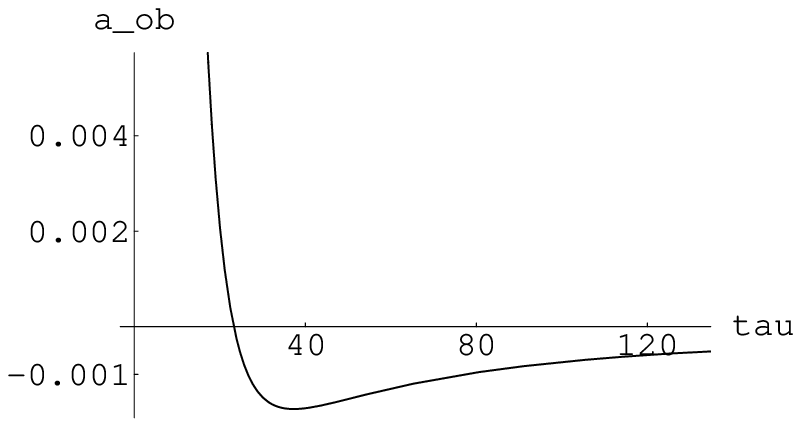}}
\caption{\label{3} The graphs of $v_{ob}$ in \eq{vt} and $a_{ob}$ in
  \eq{at} with respect to the proper time $\t$ in \eq{tt}.} 
\end{figure}

Let us consider the solution for the brane winding modes given in
\eq{c}-\eq{b}. Setting $b=1$ and choosing the negative
root\footnote{For the positive root, we find that there is no 
period of accelerated expansion.} in \eq{ab}, the metric functions
become 
\bea
A&=&-\fr{10}{3}\,\ln(\sinh t)+\fr{8}{3}\,t,\nn\\
B&=&-\fr{14}{9}\,\ln(\sinh t)+\fr{16}{9}\,t,\label{mtfunct}\\
C&=&\fr{2}{9}\,\ln(\sinh t)-\fr{4}{9}\,t,\nn
\eea
where the additive constants, which can be shifted by scaling out 
$x$ and $y$ coordinates in \eq{met1}, are set to zero. Eq. \eq{tau}
can be integrated to fix the proper time:
\be\label{tt}
\t=\fr{3}{14}\,\left[\cosh(t/3)-\sinh(t/3)\right]
\left[-21+23\,\cosh(2t)+5\,\sinh(2t)\right]\,\sinh(t)^{-7/3}.
\ee
As pointed out before, in the limits $t\to 0,\infty$ we have
$\t\to\infty,0$, respectively. Using \eq{mtfunct}, it is easy to
calculate the proper velocity and the acceleration 
of the observed space  
\bea
&&v_{ob}=\fr{dR_{ob}}{d\t}=\fr{2}{9}
e^{-8t/9}\left[-8+7\coth(t)\right]\sinh(t)^{16/9},\label{vt}\\
&&a_{ob}=\fr{d^2R_{ob}}{d\t^2}=-\fr{2}{81}e^{-32t/9}
\left[-39+88\cosh(2t)-92\sinh(2t)\right]\sinh(t)^{28/9},\label{at}
\eea
where the scale factors $R_{ob}$ and $R_{in}$ are defined in
\eq{robrin}. In figure \ref{3}, we plot $v_{ob}$ and $a_{ob}$ versus
proper time $\t$. 

From \eq{tt}, \eq{vt} and \eq{at} one finds that $v_{ob}=0$ at
$\t=\t_1\sim 6.5$ and  $a_{ob}=0$ at $\t=\t_2\sim 23.3$.
Moreover, $v_{ob}$ is positive for $\t>\t_1$ and $a_{ob}$ is positive
for $\t<\t_2$ (see figure 
\ref{3}). Therefore, in the interval $\t_1<\t<\t_2$ we have both
$v_{ob}>0$ and $a_{ob}>0$ which implies an accelerated expansion for
the observed space. The corresponding number of
e-foldings is approximately 0.7. 

It is clear that this growth is not enough for early inflation,
however the metric may explain the current observational
acceleration of the universe. In the solution, the internal space  
expands till $\t=\t_3\sim 9.9$ and then it begins to contract. It is
interesting to note that 
$\t_3\in(\t_1,\t_2)$ and thus  acceleration occurs around when the
internal space alters its evolution from expansion to
contraction. 

On the other hand, it is also possible to view the metric ``backwards in
time'' by sending $\t\to-\t$. Although this does not change the acceleration,
the velocity picks up a sign. In the reparametrized solution, we see that
as $\t$ runs from $-\t_1$ to $0$ we have $v_{ob}>0$ and
$a_{ob}>0$. This is an {\it infinite} accelerated expansion in a 
finite proper time interval and the metric ends with a singularity at
$\t=0$. As $\t\to 0^-$, the solution approaches to 
\be\label{amt1}
ds^2\,=\,-dt^2\,+\,(-\t)^{-2/3}dx^idx^i\,+\,(-\t)^{2/3}\,d\S_p^2\,,
\ee
which is the time reversal of the vacuum Kasner metric obtained in
\eq{lim1} with $\e=-1$ and in \eq{lim2} with $\e=1$. Note that the
acceleration observed in \eq{amt1} is different than the power-law
inflation and it would be interesting to calculate the spectrum of
primordial fluctuations to compare the results with the measured
spectral indices.       

Accelerating cosmologies can also be obtained with brane momentum
modes. After sending $t\to-t$, the metric functions for the solution
\eq{m1}-\eq{m2} become 
\bea
A&=&-\fr{12}{5}\,\ln(-\sinh t)+\fr{3}{5}\,t,\nn\\
B&=&-t,\label{mtfunct2}\\
C&=&-\fr{2}{5}\,\ln(-\sinh t)+\fr{3}{5}\,t,\nn
\eea
where in \eq{ab2} the positive root\footnote{It turns out that the
negative root does not yield acceleration.} is picked up
and we again set $b=1$. Note that $t$ is defined in the negative line
$(-\infty,0)$. It is possible to choose the 
integration constant in \eq{tau} so that as $t\to-\infty,0$ we have 
$\t\to-\infty,0$, respectively, and thus $\t$ is also defined in the negative
line. Calculating the proper velocity and the acceleration of
the observed space one finds 
\bea
&&v_{ob}=\fr{dR_{ob}}{d\t}=e^{-8t/5}\,\sinh(-t)^{12/5},\label{vt2}\\
&&a_{ob}=\fr{d^2R_{ob}}{d\t^2}=e^{-11t/5}\,
\left[\fr{12}{5}\cosh(-t)+\fr{8}{5}\sinh(-t)\right]\sinh(-t)^{19/5},
\label{at2} 
\eea
which are always positive and thus give an accelerated
expansion. Correspondingly, one can see that the internal
space monotonically contracts. This evolution ends with a singularity
at $\t=0$ where the metric becomes \eq{amt1} as $\t\to0^-$. The
expansion is limited near the asymptotic region $\t\to-\infty$
(i.e. as $t\to -\infty$) since  both $v_{ob}$ and $a_{ob}$
vanish. However, a large number of e-foldings can be obtained by using
the solution near $\t=0$.      

\section{Conclusions}

In this work, based on the cosmological impact of brane winding and
momentum modes, a stabilization mechanism for extra dimensions is
proposed. We find that a gas of $p$-branes wrapping the
$p$-dimensional compact space can stabilize its volume as a result of
a balance between winding and momentum pressures at the ``self-dual''
radius.  This generalizes the method studied in \cite{sb1} to $p$-branes
with $p>1$. As discussed in the introduction, a mechanism involving
$p$-branes is desirable since strings cannot keep the volume of a
generic internal manifold constant for topological reasons. On the
other hand, one should also bear in mind that it is difficult to build
late time realistic scenarios involving $p$-brane gases with $p>2$,
since the higher dimensional branes are likely to annihilate in the
early universe.  

Adding matter, we find that the fate of the stabilization depends on
the equation of state $p=\o\r$ in the observed space. For $\o>0$, the
matter can be ignored at late times and for $\o=0$ the self-dual
radius is modified but stabilization can still be achieved. When
$\o<0$, the late time dynamics is dominated by matter and one finds an
expanding internal space, which shows that the stabilization
mechanism does not work.

In this paper, we also construct accelerating solutions driven by
brane winding and momentum modes. For winding modes, a short period of
acceleration can be observed when the evolution of the internal space
changes from expansion to contraction. This suggests an interesting
mechanism for acceleration which deserves further study. Since the
number of e-foldings for the expansion is small, this metric 
cannot be used to explain the early inflation.   

In some solutions there are also accelerating periods ending with a
naked singularity, which can be seen both for winding and momentum
modes. Here, a large number of e-foldings can be obtained as one
approaches the singularity, where the solution becomes one of the
vacuum Kasner metrics. Such backgrounds may be useful in realizing
inflation in brane gas cosmology, however it is necessary to find a
way of resolving the singularity first, which may also explain the exit from
inflation. Since the internal space shrinks to zero size, the momentum
modes are expected to become important. Recall that in some solutions
the metric evolves through the singularity in the presence of
linearized momentum modes. However, the small-field approximation
should brake down near the singularity and the full momentum modes
might stop the contraction. T-duality invariance of string theory may also
play a crucial role here, so it would be interesting to include the
dilaton into the system. In any case, the quantum gravitational
effects are expected to cure the singularity, so one can try to
cut the solution and paste it to some other metric to describe the
subsequent evolution. Studying intersecting branes might also be
important in this context.

\end{document}